

\input phyzzx


\nopubblock
\line{\hfil JHU-TIPAC-930005}
\line{\hfil MSUHEP-93/02}
\line{\hfil January, 1993}
\titlepage
\title{ Inelastic Channels in $WW$ Scattering}
\author{ S.~G.~Naculich}
\address{
Department of Physics and Astronomy \break
The Johns Hopkins University  \break
Baltimore, MD  21218 }
\vskip0.6cm
\centerline{{\rm and}}
\author{ C.--P. Yuan}
\address{
Department of Physics and Astronomy \break
Michigan State University \break
East Lansing, MI  48824 }

\abstract{
If the electroweak symmetry-breaking sector becomes
strongly interacting at high energies,  it can be
probed through longitudinal $W$ scattering.  We present
a model with many inelastic channels in the $W_L W_L$
scattering process, corresponding to the production of
heavy fermion pairs.  These heavy fermions affect the
elastic scattering of $W_L$'s by propagating in loops,
greatly reducing the amplitudes in some charge channels.
We conclude that the symmetry-breaking sector cannot be
fully explored by using, for example, the $W_L^\pm W_L^\pm$
mode alone, even when no resonance is present; all
$W_L W_L \to W_L W_L$ scattering modes must be measured.
}

\endpage
\overfullrule=0pt


\def\half{{1\over 2}}

\def\tshalf{{\textstyle{1\over 2}}}
\def\d { {\rm d} }

\def\Sp{{\rm Sp}}
\def\SU{{\rm SU}}
\def\delr { \Delta r }
\def\aij { a_{J}^{~~I} }
\def\W { W_L }
\def\Z { Z_L }
\def\Lag { {\cal L} }
\def\M {{\cal M}}
\def\eps{\epsilon}
\def\ve{\varepsilon}
\def\Sig{\Sigma}
\def\Gam{\Gamma}
\def\del{\partial}
\def\gfive { \gamma_5 }
\def\Zpi{ Z_{\pi}}
\def\Sigdag{ \Sigma^\dagger }
\def\neps{{n(\eps)}}
\def\ss{\sigma_s}
\def\st{\sigma_t}
\def\mS{ m_S }
\def\mV{ m_V }
\def\GS{ \Gamma_S }
\def\GV{ \Gamma_V }
\def\pio { \pi_0 }
\def\pia { \pi_a}
\def\pib { \pi_b}
\def\piao {{ \pi_a^0 }}
\def\pibo {{ \pi_b^0 }}
\def\bpsi{{\overline \psi}}
\def\go { g_0}
\def\vo { v_0}
\def\cancel#1#2{\ooalign{$\hfil#1\mkern1mu/\hfil$\crcr$#1#2$}}
\def\slash#1{\mathpalette\cancel{#1}}
\def\pslash{ \slash{p}}
\def\delslash{ \slash{\partial} }
\def\gsim{\mathrel{\raise.3ex\hbox{$>$\kern-.75em\lower1ex\hbox{$\sim$}}}}
\def\lsim{\mathrel{\raise.3ex\hbox{$<$\kern-.75em\lower1ex\hbox{$\sim$}}}}

\chapter {Introduction}

\REF\ETone{
J. M. Cornwall, D. N. Levin, and G. Tiktopoulos,
{\sl Phys. Rev.} {\bf D10} (1974)  1145; \nextline
C.~Vayonakis,
{\sl Lett. Nuovo Cimento } {\bf 17} (1976) 383; \nextline
B.~W.~Lee, C.~Quigg, and H.~Thacker,
{\sl Phys. Rev.} {\bf D16} (1977)  1519.}
\REF\ETtwo{
M.~S.~Chanowitz and M.~K.~Gaillard,
{\sl Nucl. Phys.} {\bf B261} (1985)  379; \nextline
G.~J.~Gounaris, R.~Kogerler, and H.~Neufeld,
{\sl Phys. Rev.} {\bf D34} (1986)  3257.}
\REF\ETthree{
Y.--P. Yao and C.--P. Yuan,
{\sl Phys. Rev.} {\bf D38} (1988)  2237; \nextline
J. Bagger and C. R. Schmidt,
{\sl Phys. Rev.} {\bf D41} (1990)  264; \nextline
H. Veltman,
{\sl Phys. Rev.} {\bf D41} (1990)  2294; \nextline
H.--J. He, Y.--P. Kuang, and X. Li,
{\sl Phys. Rev. Lett.} {\bf 69} (1992)  2619.}
\REF\rrNolose{
M. S. Chanowitz and M. K. Gaillard,
{\sl Nucl. Phys.} {\bf B261} (1985) 379;\nextline
M. S. Chanowitz,
{\sl Ann. Rev. Nucl. Part. Sci.} {\bf 38} (1988) 323.}

\REF\rrWpwp{
M.~S.~Chanowitz and M.~Golden,
{\sl Phys.~Rev.~Lett.} {\bf 61} (1988) 1053;
{\bf 63} (1989) 466(E); \nextline
M.~S.~Berger and M.~S.~Chanowitz,
{\sl  Phys. Lett.} {\bf B263} (1991) 509.}

\REF\rrInel{
R. S. Chivukula and M. Golden,
{\sl Phys. Lett.} {\bf B267} (1991) 233; \nextline
S.~G.~Naculich and C.-P.~Yuan,
{\sl Phys. Lett.} {\bf B293} (1992) 395; \nextline
R. S. Chivukula, M. Golden, D. Kominis, and M. V. Ramana,
{\sl Phys. Lett.} {\bf B293} (1992) 400; \nextline
D. Morris, R. Peccei, and R. Rosenfeld,
preprint UCLA-92-TEP-45.}
\REF\rrNY{
S.~G.~Naculich and C.--P.~Yuan,
{\sl Phys. Lett.} {\bf B293} (1992) 405.}

One of the major goals of the SSC is to explore the physics
which spontaneously breaks the electroweak symmetry of
the standard model.
This symmetry breaking gives rise to Goldstone bosons,
which become the longitudinal components $\W$
of the massive vector gauge bosons.
(We use $W$ to denote either the $W^\pm$ or $Z^0$ boson.)
Consequently, the symmetry-breaking sector can be directly
probed by scattering longitudinal $W$'s.

At energies $s \gg M_W^2$,
the $\W \W$ scattering amplitudes,
by virtue of the equivalence theorem,
become approximately equal
to the scattering amplitudes of the Goldstone bosons
of the broken symmetry [\ETone--\ETthree].
If we assume the existence of a custodial SU(2) symmetry,
the pattern of symmetry breaking is
$ \SU(2)_L \times  \SU(2)_R \to \SU(2)_V$
and the Goldstone boson scattering amplitudes
are given by
$$
\eqalign{
\M (\Z^0 \Z^0 \to \W^- \W^+) & = A(s,t,u), \cr
\M (\W^- \W^+ \to \Z^0 \Z^0) & = A(s,t,u), \cr
\M (\W^- \W^+ \to \W^- \W^+) & = A(s,t,u)+A(t,s,u), \cr
\M (\Z^0 \Z^0 \to \Z^0 \Z^0) & = A(s,t,u)+A(t,s,u)+A(u,t,s), \cr
\M (\W^\pm \Z^0 \to \W^\pm \Z^0) & = A(t,s,u), \cr
\M (\W^\pm \W^\pm \to \W^\pm \W^\pm) & =A(t,s,u)+A(u,t,s). \cr}
\eqn\eescattamp
$$
In an energy expansion of the amplitudes,
the lowest--order term is uniquely determined to be
$$
A(s,t,u) = {s \over  f^2} + \cdots,
\eqn\eeLET
$$
where $f=250$ GeV is the scale of symmetry-breaking.

Higher--order terms in the energy expansion of the amplitude
\eeLET~are sensitive to the specific form of the
symmetry breaking mechanism.
For example,
if there is a light Higgs boson,
the next term in the expansion is
$ s^2 / M_H^2 f^2 $,
signalling the approach to a resonance
in the $\W \W $ scattering amplitudes
at $ s \sim M_H^2$.
On the other hand,
if there are no light resonances,
the low--energy  result \eeLET~will
continue to hold at higher energies.
This amplitude grows with increasing center--of--mass
energy, becoming strong at around 1 TeV.
(Of course, corrections to eq.~\eeLET~must
eventually become important,
because the lowest--order term violates partial--wave unitarity,
\ie,
$ |{\rm Re}~a_0^0|  > \half $,
at about  $ 2 \sqrt{2\pi} f \sim 1.3$ TeV.)
This enhanced $\W \W$ scattering amplitude
indicates the presence of new strong interactions
at or above 1 TeV.
It has been claimed that the energy
and luminosity of the SSC are large enough that,
whatever form the new interactions take,
they would be observable in $\W \W $
two--body interactions via leptonic decays of $\W$'s [\rrNolose].
In particular, the like-sign mode $\W^\pm \W^\pm$ is a favorite
candidate for observing these new strong interactions
because the standard model background for this mode is small [\rrWpwp].
This prediction of strong $\W \W$ scattering
in the absence of a light resonance is called the
``no--lose theorem.''

In this paper, we explore the possibility
that there are inelastic channels in the $\W \W$
scattering process [\rrInel,\rrNY],
and the implications of these
inelastic processes for the no-lose theorem.
In this scenario,
the $\W$'s would scatter
not only into a $\W \W$ final state
but also into other final states.
Even if the production of these other particles
is not directly observed,
they would affect the {\it elastic} scattering of $\W$'s by
propagating in loops.
These loops necessarily contribute to the
imaginary part of the elastic scattering amplitude,
which is related by the optical theorem to the total cross-section.
The loops also contribute to
the real part of the elastic amplitude,
interfering with the Born contribution.

If the number of inelastic channels is large,
these loop effects will be strong,
and may significantly
reduce the signal in some charge channels,
{\it e.g.} the $\W^\pm \W^\pm$ channels,
making it difficult to observe.
To exemplify this possibility,
we examine in this paper a model
in which the inelastic channels take the form
of heavy fermions,
which are pair--produced in $\W \W$ scattering.
Unlike the case of a Higgs boson,
production of these particles
will not produce resonances
in the elastic $\W \W $ scattering amplitudes.
They alter the amplitudes
through loop effects, however,
and lead to behavior markedly differently from
the low--energy result \eeLET~at scales above
the threshold for fermion pair production
but below 1 TeV.
(New physics must enter at around 1 TeV,
because the amplitudes violate unitarity
not far above this scale.)
The lesson is that
to be certain of detecting the symmetry--breaking sector
it will be necessary to measure scattering
in all the final state $\W \W $ modes.

\chapter {The Large-$N$ Fermion Model}

The scattering of longitudinal $W$ bosons
can be studied
within the framework of a nonlinear sigma model.
This is a minimalist approach
because the model contains only
the Goldstone bosons $\pi$
of the spontaneously broken symmetry,
parametrized by the matrix
$$
 \Sigma
=\exp{\left( i \tau_a \pia  \over f \right) },
 \quad\quad  a=1,2,3,
\eqn\eeSigma
$$
where $\tau_a$ are the Pauli matrices.
By the equivalence theorem,
the Goldstone bosons $\pi$
correspond to the longitudinal degrees of freedom $\W$
of the vector gauge bosons.

To study the effects of inelastic channels
in $\W \W$ scattering,
we couple the Goldstone bosons
to $N$ degenerate fermion doublets
$\psi^j$
with mass $m=gv$.
The effects of these fermions
will be most important when
the Yukawa coupling $g$ is large.
To capture this,
we will not calculate the amplitudes
perturbatively in $g$,
but rather in a $1/N$ expansion.
The results will be valid for arbitrary Yukawa coupling $g$,
\ie, for all values of the fermion mass $m$.
(Were we to calculate perturbatively in the Yukawa coupling,
the real part of the loop correction would contribute
through interference with the tree--level amplitude,
but the imaginary part would be higher order.
In the large--$N$ approach,
the imaginary part
of the loop correction
contributes in leading order,
and will be important above the
threshold for production of fermions.)

The Lagrangian for our model [\rrNY] is
$$
\Lag
=
{N \vo^2 \over 4} \,
{\rm Tr} \left(\del_\mu \Sigma \del^\mu \Sigdag \right)
+ \sum^{N}_{j=1} \bigg[ \bpsi^j  ~i\delslash ~ \psi^j
- \go \vo \left( \bpsi_L^j \Sigma  \psi_R^j
          + \bpsi_R^j \Sigdag \psi_L^j \right) \bigg],
\eqn\eeNLSMLag
$$
with
$$
 \Sigma
=\exp{\left( i \tau_a \piao \over \sqrt{N} \vo \right) },
 \quad\quad
\psi_L^j  = \tshalf \left( 1 - \gamma_5 \right) \psi^j ,
 \quad\quad
\psi_R^j  = \tshalf \left( 1 + \gamma_5 \right) \psi^j .
$$
The Feynman rules are obtained by
expanding $\Sigma$ in terms of the Goldstone fields $\piao$.
The first term in the Lagrangian \eeNLSMLag~yields
vertices involving only Goldstone bosons:
$$
 {\rm Tr} \left(\del_\mu \Sigma \del^\mu \Sigdag \right)
 = {2 \over N \vo^2} ( \del_\mu \piao ) ( \del^\mu \piao )
 + {2 \over 3 N^2 \vo^4}
  \Bigl[ \piao (\del_\mu \piao) \pibo (\del^\mu \pibo)
       - \piao \piao (\del_\mu \pibo) (\del^\mu \pibo) \Bigr]
 + \cdots
\eqn\eeGoldvert
$$
The last term in eq. \eeNLSMLag~yields
fermion--Goldstone boson vertices:
$$
\bpsi_L^j \Sigma  \psi_R^j  + \bpsi_R^j \Sigdag \psi_L^j
= \bpsi^j \left[
1
+  { i \tau_a \piao \over    N^{1/2} \vo } \gfive
-  { \piao \piao \over 2  N       \vo^2}
-  { i (\tau_a \piao) (\pibo \pibo) \over 6  N^{3/2} \vo^3} \gfive
-  { (\piao \piao) (\pibo \pibo) \over 24 N^2     \vo^4}
+ \cdots  \right] \psi^j.
\eqn\eefermvert
$$
We will use these vertices to compute Green functions
to leading order in $1/N$,
holding the parameters
$\go$ and $\vo$ fixed as $N \to \infty$.

The Lagrangian \eeNLSMLag~is {\it not} the most general one
with global $ \SU(2)_L \times  \SU(2)_R$ chiral symmetry;
we have omitted a possible derivative coupling of the form
$  \kappa_L \bpsi_L  (\Sigma i \delslash  \Sigdag) \psi_L
+  \kappa_R \bpsi_R  (\Sigdag i \delslash  \Sigma) \psi_R $.
(If parity is conserved, then $\kappa_L = \kappa_R$.)
We also have not included any four-derivative terms
involving  $\Sigma$.
To leading order in $1/N$,
no such terms are needed to
absorb divergences;
all divergences due to fermion loops
can be absorbed into the bare parameters
in \eeNLSMLag.

To leading order in $1/N$,
the only corrections to the Green functions
come from fermion loops.
There are no leading order corrections
to the fermion self energy,
so the inverse fermion propagator remains that
of a free field,
$$
\Gamma^{(2)}_{\psi \bpsi} (p) = \pslash - m,
\eqn\eefermtwopt
$$
with mass
$$
m = \go \vo.
\eqn\eefermmass
$$
There is no fermion wavefunction renormalization
to leading order in $1/N$.

Fermion loops contribute a divergence to the
Goldstone boson self energy.
Regularizing the fermion loop integral in $d= (4-\eps)$ dimensions,
we find
$$
  \Gamma^{(2)}_{\piao \pibo } (p^2)
    =
     \left[\left( 1 + {\go^2 \neps \over 2\pi^2 \eps m^{\eps} }
                 - { \go^2 \over 4\pi^2 }
           \right) p^2 + O(p^4)
     \right] \delta_{ab},
\eqn\eeGoldtwopt
$$
where
$\neps  = 1 + \tshalf \eps ( 1 - \gamma + \ln 4\pi )$
and
$\gamma = 0.577215 \cdots.$
The two-point function \eeGoldtwopt~vanishes at $p^2 = 0$;
the Goldstone boson remains massless.
The Goldstone boson wavefunction renormalization
$$
\piao    = \Zpi^{1/2}   \pia,
\qquad
\vo      = \Zpi^{1/2}   v,
\qquad
\Gamma^{(n)}_{\pio}
         = \Zpi^{-n/2} \Gamma^{(n)}_{\pi}
\eqn\eeGoldwavefcn
$$
is chosen to be
$$
      \Zpi^{-1}
  = 1 + {\go^2 \neps \over 2\pi^2 \eps m^{\eps} } - {\go^2 \over 4\pi^2 },
\eqn\eeZpiinv
$$
so that the renormalized Goldstone propagator
has unit on-shell residue
$$
{\d \Gam^{(2)}_{\pia  \pib} \over \d p^2 } \bigg|_{p^2 = 0} = \delta_{ab}.
\eqn\eeGoldres
$$
We can rewrite eq.~\eeZpiinv~as
$$
\Zpi
  = 1 - {m^{2-\eps} \neps \over 2\pi^2 v^2 \eps } +  { m^2 \over 4\pi^2 v^2}
\eqn\eeZpi
$$
using
$ \Zpi^{-1}  = v^2/\vo^2$
and
$ \go = m/\vo $.

\REF\rrSpence{
G. 't Hooft and M. Veltman,
{\sl Nucl.~Phys.} {\bf B153} (1979)  365.}
The wavefunction renormalization \eeGoldwavefcn~is
the only renormalization necessary in this model
to leading order in $1/N$.
In particular, the divergences in the
Goldstone boson four--point functions \eescattamp~due
to fermion loops are precisely absorbed by
the Goldstone boson wavefunction renormalization.
(If Goldstone boson loops were not suppressed
by a factor of $1/N$,
additional four-derivative counterterms would be needed.)
Adding up all the fermion loop contributions to the
four-point functions,
we obtain the renormalized amplitude
$$
A(s,t,u)
 = {1\over N} \bigg\{
{s\over v^2} - {m^2 \over 4 \pi^2 v^4} s F_2 (s)
- {m^4 \over 4 \pi^2 v^4}
\Big[ F_4 (s,t)  + F_4 (s,u)  -  F_4 (t,u)  \Big] \bigg\},
\eqn\eeNLSMamp
$$
where the functions $F_2 (s)$ and $F_4 (s,t)$ are defined by
$$
\eqalign{
F_2 (s)
& = \int_0^1 \d x
\ln \left[ 1 - {s\over m^2} x(1-x) - i \ve \right],
\cr
F_4 (s,t)
&
= \int_0^1  \d x
\left[ x^2 - x + { m^2(s+t)\over st } \right]^{-1}
\cr
&
{}~~~~\times \biggl\{ \ln \Bigl[ 1 - {s\over m^2} x(1-x) - i \ve \Bigr]
+       \ln \Bigl[ 1 - {t\over m^2} x(1-x) - i \ve \Bigr]
\biggr\}.
\cr}
\eqn\eeFintegrals
$$
The integral $F_2(s)$ is given by
$$
\eqalign{
F_2(s<0)
&  =  -~2+~\sqrt{ 1 - {4m^2 \over s} }
  ~\ln\left(
  { \sqrt{4m^2-s} + \sqrt{-s}  \over
    \sqrt{4m^2-s} - \sqrt{-s}  } \right) ,
\cr
F_2( 0<s<4m^2)
& =-~2+2~\sqrt{ -1 + {4m^2 \over s}  } ~{\rm arctan}
  \sqrt{ {s \over 4m^2-s} },
\cr
F_2(s>4m^2)
& =-~2+\sqrt{ 1 - {4m^2 \over s}}
\left[ \ln\left(
 { \sqrt{s} + \sqrt{s-4m^2} \over   \sqrt{s} -  \sqrt{s-4m^2} }
\right) - i  \pi\right],
\cr}
\eqn\eeFtwo
$$
and the integral $F_4 (s,t)$ can be written [\rrSpence]
in terms of Spence functions as
$$
\eqalign{
F_4&(s,t)
= 2 \left[ 1 - { 4m^2(s+t)\over st } \right]^{-\half}
\bigg[ \Sp \left( {x_+  \over x_+ - y_+(s) } + i\ss\ve \right)
\cr
&
     + \Sp \left( {x_+  \over x_+ - y_-(s) } - i\ss\ve \right)
     - \Sp \left( {-x_- \over x_+ - y_+(s) } + i\ss\ve \right)
     - \Sp \left( {-x_- \over x_+ - y_-(s) } - i\ss\ve \right)
\cr
&
     + \Sp \left( {x_+  \over x_+ - y_+(t) } + i\st\ve \right)
     + \Sp \left( {x_+  \over x_+ - y_-(t) } - i\st\ve \right)
     - \Sp \left( {-x_- \over x_+ - y_+(t) } + i\st\ve \right)
\cr
&
     ~~~~~~~~~~~~~~~~~~~~~
     - \Sp \left( {-x_- \over x_+ - y_-(t) } - i\st\ve \right)
     + 2\pi i \Theta( st ) \ln \left( x_+ \over - x_- \right)
     \bigg],
\cr}
\eqn\eeFfour
$$
where
$$
x_\pm = \half \pm \sqrt{ {1\over 4} - {m^2 (s+t)\over st} } ~,
\qquad
y_\pm (s) = \half \pm \sqrt{ {1\over 4} - {m^2 \over s} } ~,
\eqn\eexpm
$$
and  $\ss$ denotes the sign of $s$.

Now we fix $N$ to a finite value,
the number of fermion doublets in the model.
Then we set the scale $v$ equal to $f/\sqrt{N}$,
where $f = 250$ GeV is the scale of symmetry breaking,
to obtain
$$
A(s,t,u)
=   {s\over f^2}
  - {N m^2 \over 4 \pi^2 f^4} s F_2 (s)
  - {N m^4 \over 4 \pi^2 f^4}
\Big[ F_4 (s,t)  + F_4 (s,u)  -  F_4 (t,u)  \Big].
\eqn\eeNLSMampf
$$
In the limit $s \ll m^2$, the amplitude
$ A(s,t,u) $ approaches $ s/f^2 $,
in accord with the low--energy  result \eeLET.
The $\W \W$ scattering amplitudes
are obtain by substituting eq.~\eeNLSMampf~into
eq.~\eescattamp.
The partial waves are defined by
$$
\aij = {1\over 64\pi} \int_{-1}^{1} \d (\cos\theta)
{}~P_J (\cos \theta) ~T(I),
\eqn\eepartial
$$
where the isospin amplitudes $T(I)$ are given by
$$
\eqalign{
T(0)
&
= 3 A(s,t,u) + A(t,s,u) + A(u,t,s),
\cr
T(1)
&
= A(t,s,u) - A(u,t,s),
\cr
T(2)
&
= A(t,s,u) + A(u,t,s).
\cr
}
\eqn\eeisoamp
$$

The model we have just presented
depends on only two parameters:
the number of degenerate fermion doublets $N$
and the fermion mass $m$.
In sect. 3,
we will examine the behavior
of this model in the heavy fermion limit.
In sect. 4,
we will consider the model with
a large number of somewhat lighter fermions ($m \sim 250$ GeV),
in order to study the effect of inelastic channels
on elastic $\W \W$ scattering in the TeV region.

\chapter{The heavy fermion limit}

In this section,
we will examine the effect
of very heavy fermions on elastic $\W \W$ scattering
at energies {\it below} the threshold for fermion pair production.
The heavy fermion limit is obtained by letting the
Yukawa coupling become large, holding the symmetry
breaking scale fixed at $f=250$ GeV.
This procedure is legitimate in our model,
since we have calculated the scattering amplitudes
non-perturbatively in the Yukawa coupling
(to leading order in $1/N$).

We consider a model with $N=3$
species of degenerate fermion doublets
of mass $m = 1$ TeV.
This corresponds to 1 additional generation
of quark doublets,
due to the color degeneracy.
The non-vanishing
$J=0$ and $J=1$
partial waves of
the elastic $\W \W$ scattering amplitudes
are shown by the solid lines in fig.~1.
These are to be compared with the partial waves
of the low--energy  amplitude $A(s,t,u) = s/f^2$,
which are shown by the dotted lines.
We note that
the isospin $I=0$ and $I=1$ partial waves
increase somewhat faster with energy
due to the heavy fermions,
violating unitarity at a lower scale.
For $N=3$,
the $a_0^0$ partial wave
exceeds $\half$ at around 1.2 TeV
(with $|a_0^0| < 1$  up to about 1.6 TeV);
unitarity is violated at an even lower scale for larger $N$.
On the other hand, the isospin $I=2$ partial wave,
which alone contributes to the
$\W^\pm \W^\pm$ scattering amplitude,
is greatly reduced at high energies by
heavy fermion loop effects.

The effect of taking the fermion mass even
larger can be seen by noting that
the amplitude \eeNLSMampf~has a well defined limit
as $m \to \infty$, namely
$$
A(s,t,u)
{}~\to~{s\over f^2}
  ~-~ {N \over 48 \pi^2} {s^2 \over f^4}
  ~+~ {N \over 48 \pi^2} {t^2 + u^2 \over f^4} ,
\qquad \qquad
m \to \infty.
\eqn\eeamplimit
$$
The fermions do not completely decouple
as their mass increases,
but leave behind mass-independent contributions
of $O(s^2)$.
The partial waves of the amplitudes
in this infinite mass limit
(with $N=3$)
are shown in fig.~1 by the dot-dashed lines.

\REF\rrGL{
J. Gasser and H. Leutwyler,
{\sl Ann. Phys.} {\bf 158} (1984) 142;
{\sl Nucl. Phys.} {\bf B250} (1985) 465.}
\REF\rrBag{J. Bagger, in
{\it Perspectives in the Standard Model},
eds. R. Ellis, C. Hill, and J. Lykken
(World Scientific, Singapore, 1992).}
\REF\rrAit{
I. Aitchison,
{\sl Acta Phys. Polon.} {\bf B18} (1987) 191, and references therein.}
\REF\rrDV{
S. Dawson and G. Valencia,
{\sl Nucl. Phys.} {\bf B348} (1991) 23.}
\REF\rrTru{
T. N. Truong,
{\sl Phys. Lett.} {\bf B273} (1991) 292.}

The $O(s^2)$ contributions to the scattering amplitudes
\eeamplimit~correspond to four-derivative terms
in the low-energy effective Lagrangian [\rrGL, \rrBag]
$$
\Lag_4
= {L_1 \over 16\pi^2}
  \Tr \left(\del_\mu \Sigdag \del^\mu \Sig\right)
  \Tr \left(\del_\nu \Sigdag \del^\nu \Sig\right)
+ {L_2 \over 16\pi^2}
  \Tr \left(\del_\mu \Sigdag \del_\nu \Sig\right)
  \Tr \left(\del^\mu \Sigdag \del^\nu \Sig\right) .
\eqn\eefourderiv
$$
Very heavy fermions induce four-derivative
terms with coefficients [\rrAit, \rrDV]
$$
L_1 = -~{N\over 24},
\qquad\qquad
L_2 = {N\over 12}.
\eqn\eefermcoeff
$$
These four-derivative terms are a linear combination
of those induced by an isoscalar  spin--zero resonance
and those induced by an isovector  spin--one resonance [\rrAit, \rrTru].
A heavy scalar particle with mass $\mS$
and width $\GS = 3 g^2 \mS^3 / 32 \pi f^2$
induces four-derivative terms with coefficients
$$
L_1 = {64 \pi^3 f^4 \GS \over 3 \mS^5},
\qquad\qquad
L_2 = 0.
\eqn\eescalarcoeff
$$
(A standard model Higgs boson corresponds to $g=1$.)
A heavy vector meson with mass $\mV$
and width $\GV$
induces four-derivative terms with coefficients
$$
L_1 = - ~{192 \pi^3 f^4 \GV \over \mV^5},
\qquad\qquad
L_2 = {192 \pi^3 f^4 \GV \over \mV^5}.
\eqn\eevectorcoeff
$$
Thus $N$ heavy fermion doublets induce $O(s^2)$ terms
in the scattering amplitude
equivalent to those induced by
the combination of a scalar particle with
$\mS^5  = 512 \pi^3 f^4 \GS / N $
and a vector particle with
$\mV^5 = 2304 \pi^3 f^4 \GV / N $.
In the model with 3 degenerate very heavy fermion doublets,
the four--derivative terms are equivalent
to those of a scalar resonance
with $\mS \sim \GS \sim 2$ TeV
together with a scaled--QCD $\rho$
with $\mV \sim 2$ TeV and $\GV \sim 0.4$ TeV.
At an energy scale much less than the masses of the fermions
and the scalar and vector resonances,
it is not possible to
differentiate these two symmetry--breaking mechanisms.
Thus, it is important to probe longitudinal $W$ interactions
in the TeV region.

\chapter{ Inelastic channels in $\W \W$ scattering}

The primary goal of this paper
is to examine the effect of inelastic channels
on elastic $\W \W$ scattering.
In sect.~2,
we presented a model
containing additional species of heavy fermions,
which can be produced in $\W \W$ scattering.
The $\W \W$ scattering only becomes inelastic
above the threshold for fermion pair production,
however, so we do not want the fermions to be too heavy.
In this section, we will choose the fermion mass to be
$m=250$ GeV
so that $\W \W$ scattering in this model will be inelastic
in an energy range accessible to the SSC
($\sqrt{S} = 40$ TeV).

\REF\rrWmass{
M. Veltman,
{\sl Phys. Lett.} {\bf 91B} (1980) 95;\nextline
W. J. Marciano and A. Sirlin,
{\sl Phys. Rev.} {\bf D22 } (1980) 2695;\nextline
S. Bertollini and A. Sirlin,
{\sl Nucl. Phys.} {\bf B248} (1984) 589.}
\REF\rrSparam{
M. Peskin and T. Takeuchi,
{\sl Phys. Rev. Lett.} {\bf 65} (1990) 964.
}
\REF\rrMarc{
W. Marciano,
``Precision Tests of Electroweak Theory,''
preprint BNL-47278.}
The greater the number of additional fermion species,
the more pronounced will be their effect on elastic $\W \W$ scattering.
There is an experimental constraint on the number $N$
of additional fermion doublets, however,
due to their contribution to $\delr$ [\rrWmass],
or equivalently, to the $S$ parameter [\rrSparam].
For $N$ degenerate heavy fermion doublets,
the shift in $\delr $ is
$$
\delr = {\alpha N \over 12 \pi \sin^2 \theta_W},
\eqn\eone
$$
where $\theta_W$ is the weak mixing angle defined in the
$(\alpha,\, G_F, \, M_Z)$ scheme.
Equivalently,
the shift in the $S$ parameter is [\rrMarc]
$$
S = {N \over 6\pi} .
\eqn\etwo
$$
The bounds
$\delr \lsim 0.015$
or
$S \lsim 1$
require $N \lsim 16$.
In this section, we will choose $N=15$,
corresponding to 5 additional generations
of degenerate quark doublets,
due to the color degeneracy.

In fig.~2,
we show the non-vanishing
$J=0$ and $J=1$
partial waves of
the elastic $\W \W$ scattering amplitudes
for the model with $N=15$ and $m=250$ GeV.
This figure also shows the partial waves for the low--energy  result
$A (s,t,u) = s/f^2$.
The magnitudes of the $a_0^0$ partial waves
in the model with 15 heavy fermion doublets
and in the low--energy model
are similar,
and both remain unitary ($ |a_0^0| < 1$)  below 1.5 TeV.
However,
the heavy fermions greatly suppress the
$ a_1^1$ and $ a_0^2 $ partial waves
relative to the low--energy  result.

In fig.~3,
we show the constituent cross-sections
for $\W \W$ scattering.
The contributions to the cross-sections due to the real
(solid lines) and imaginary (dashed lines) parts of the amplitudes
are plotted separately.
The total cross-section is given by the sum of the two contributions.
We also show (dotted lines) the cross-sections for the low--energy
result.

\REF\rrBeam{ G. A. Kane, S. Mrenna, S. G. Naculich, and C.--P. Yuan,
in preparation.}
We see from fig.~3 that the cross-sections for scattering into
the $\W^+ \W^-$  and $\Z^0 \Z^0$ modes
are somewhat increased in the heavy fermion model
relative to the low--energy  result,
because of enhancement from the loop contribution
to the scattering amplitudes.
The cross-sections for the $\W^+ \W^-$ and $\Z^0 \Z^0$ modes
in the TeV region are almost entirely due to
the imaginary forward part of the amplitude,
the effect becoming stronger for higher $M_{WW}$.
The imaginary forward part of the amplitude
is related by the optical theorem to the total cross-section,
which is large due to the presence of many inelastic channels [\rrBeam].

The cross-sections for the $\W^+ \Z^0$ and $\W^+ \W^+$ modes,
however,
are greatly reduced in the heavy fermion model
relative to the low--energy result.
The $\W^+ \W^+$ cross-section at $M_{WW}=1.5$ TeV
is down by about a factor of 10
from the low-energy result,
and the $\W^+ \Z^0$ mode is reduced by a similar factor.
This occurs because
the tree level amplitude is largely cancelled
by the real part of the loop amplitude,
and the imaginary part of the amplitude
is zero for $\W^+ \W^+$
and small for $\W^+ \Z^0$.

\REF\rrPhen{The phenomenology of the low--energy model
is discussed in
J. Bagger, V. Barger,  K. Cheung, J. Gunion,
T. Han, G. Ladinsky, R. Rosenfeld, and C.--P. Yuan,
in preparation.}
\REF\rrWkt{
J. G. Morfin and Wu--Ki Tung,
{\sl Z. Phys.} {\bf C52} (1991)  13.}
The event rates for scattering processes at the SSC
are obtained by folding these constituent cross-sections with
the parton luminosities.
Our purpose here is not to study the phenomenology of
the heavy fermion model in detail,
but to illustrate the differences
between this model and the low--energy model [\rrPhen].
In fig.~4,
we show the production rates of longitudinal $W$ pairs
in one SSC year (with $10^4$ pb$^{-1}$)
for various charge modes
as a function of the invariant mass $M_{WW}$.
As we anticipated,
the $\W^+ \W^+$ and $\W^+ \Z^0$ event rates
in the heavy fermion model are greatly reduced
in the TeV region.
The $\W^+ \W^-$ and $\Z^0 \Z^0$ event rates
are increased,
but by less than a factor of two
for $M_{WW} \lsim 1.5$ TeV.
In fig.~4, the branching ratio
of the $W$ boson decay has not been included.
Each decay product of the $W$'s
is required to have a minimum transverse momentum of
20 GeV within rapidity $\pm 2.5$.
The event rate is calculated using the effective--$W$ approximation.
The parton distribution function used is the leading order set,
Fit SL, of Morfin and Tung [\rrWkt].
The scale used in evaluating the parton distribution function
in conjunction with the effective--$W$ method is $M_W$.

As seen from fig.~2,
the $a_0^0$ partial wave of the heavy fermion model
does not differ significantly
from that of the low--energy model
up to the scale ($\sim 1.5$ TeV)
at which they both violate unitarity.
Beyond 1.5  TeV,
$|a_0^0|$ exceeds 1,
at which point
new physics must enter to unitarize the physical amplitudes.
In the literature,
various methods, such as saturating or unitarizing the partial waves,
have been used to suggest what might happen when
the partial waves of the low--energy model
violate perturbative unitarity.
Using these methods,
we expect the $a_0^0$ partial waves
to have roughly the same behavior
in both the heavy fermion model and the
low--energy model.
However, the $a_1^1$ and $a_0^2$ partial waves in
the heavy fermion model
remain small relative to those in the low--energy model.
This implies that,
for $M_{WW} \lsim 3$ TeV,
the $\W^\pm \W^\pm$ and $\W^\pm \Z^0$
event rates predicted by this model
will always be much smaller than the
corresponding event rates in the low--energy model.

\chapter{Concluding Remarks}

In this paper,
we have examined the effects
of inelastic channels
in the $\W \W $ scattering process
in one specific model.
In this model,
containing heavy fermion doublets,
the rate of elastic $\W \W $ scattering
at energies above the threshold for fermion pair production
differs significantly
from the low--energy result.
In particular,
we found a large suppression
of the $\W^\pm  \W^\pm $ mode.
This implies that,
even when the symmetry--breaking sector does
not contain any resonances,
the $\W^\pm  \W^\pm  \to \W^\pm  \W^\pm $ interaction
does not necessarily become strong
as it does in the low--energy model.
A lesson to be drawn from this model
is that all charge modes of the $\W \W \to \W \W $ process
need to be observed
to be sure of detecting the symmetry--breaking sector.

\endpage

\centerline{\fourteenrm Acknowledgments }

It is a pleasure to thank G.~L.~Kane
for asking the questions which stimulated this work,
and for many fruitful discussions.
We are also grateful to
J.~Bagger,
J.~Bjorken,
Gordon Feldman,
B.~Grinstein,
C.~Im,
G.~Ladinsky,
S.~Meshkov,
S.~Mrenna,
F.~Paige,
and
E.~Poppitz
for discussions.
The work of S.G.N. has been supported by the National Science Foundation
under grant no.~PHY-90-96198.
The work of C.P.Y. was funded in part by TNRLC grant no. RGFY9240.

\FIG\fig{  The partial waves $\aij$
with isospin $I$ and angular momentum $J$
of the elastic $\W \W$ scattering amplitudes
for 3 fermion doublets with $m=1$ TeV (solid lines),
with $m=\infty$ (dot-dashed lines),
and in the absence of heavy fermions (dotted lines). }

\FIG\fig{  The partial waves $\aij$ of the
elastic $\W \W$ scattering amplitudes
for 15 fermion doublets with $m=250$ GeV
(solid lines = real part, dashed lines = imaginary part),
and in the absence of heavy fermions
(dotted lines = real part, no imaginary part). }

\FIG\fig{  Constituent cross-sections
in the model with
15 fermion doublets of mass 250 GeV.
The solid (dashed) lines show the contributions
due to the real (imaginary) part of the amplitude.
The dotted lines show the constituent cross-sections
in the model with no heavy fermions. }

\FIG\fig{  Event rates
in the model with
15 fermion doublets of mass 250 GeV.
The solid lines show the invariant mass $M_{WW}$ distributions
 of this model.
The dotted lines show the event rates
in the model with no heavy fermions. }

\refout

\figout

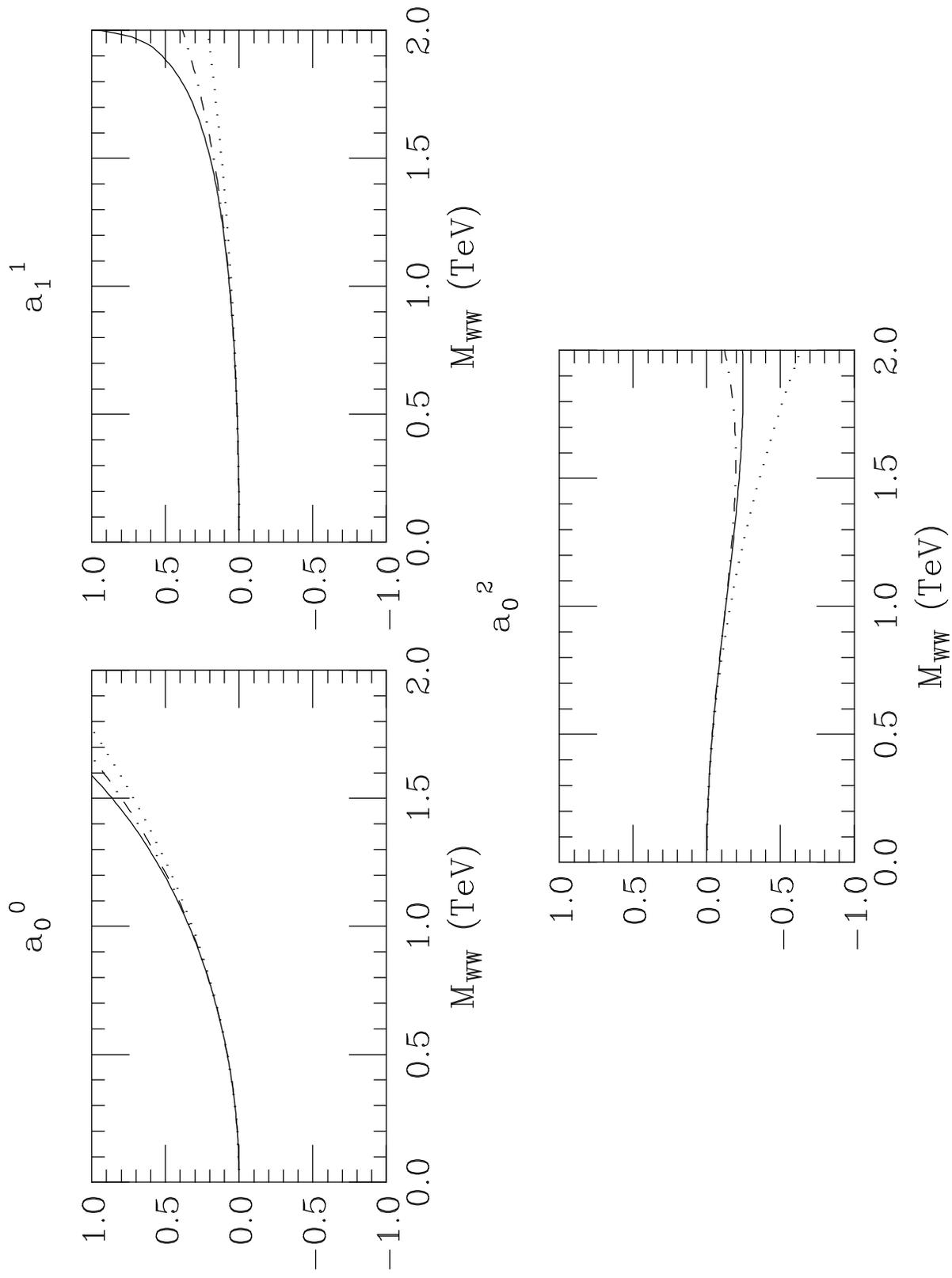
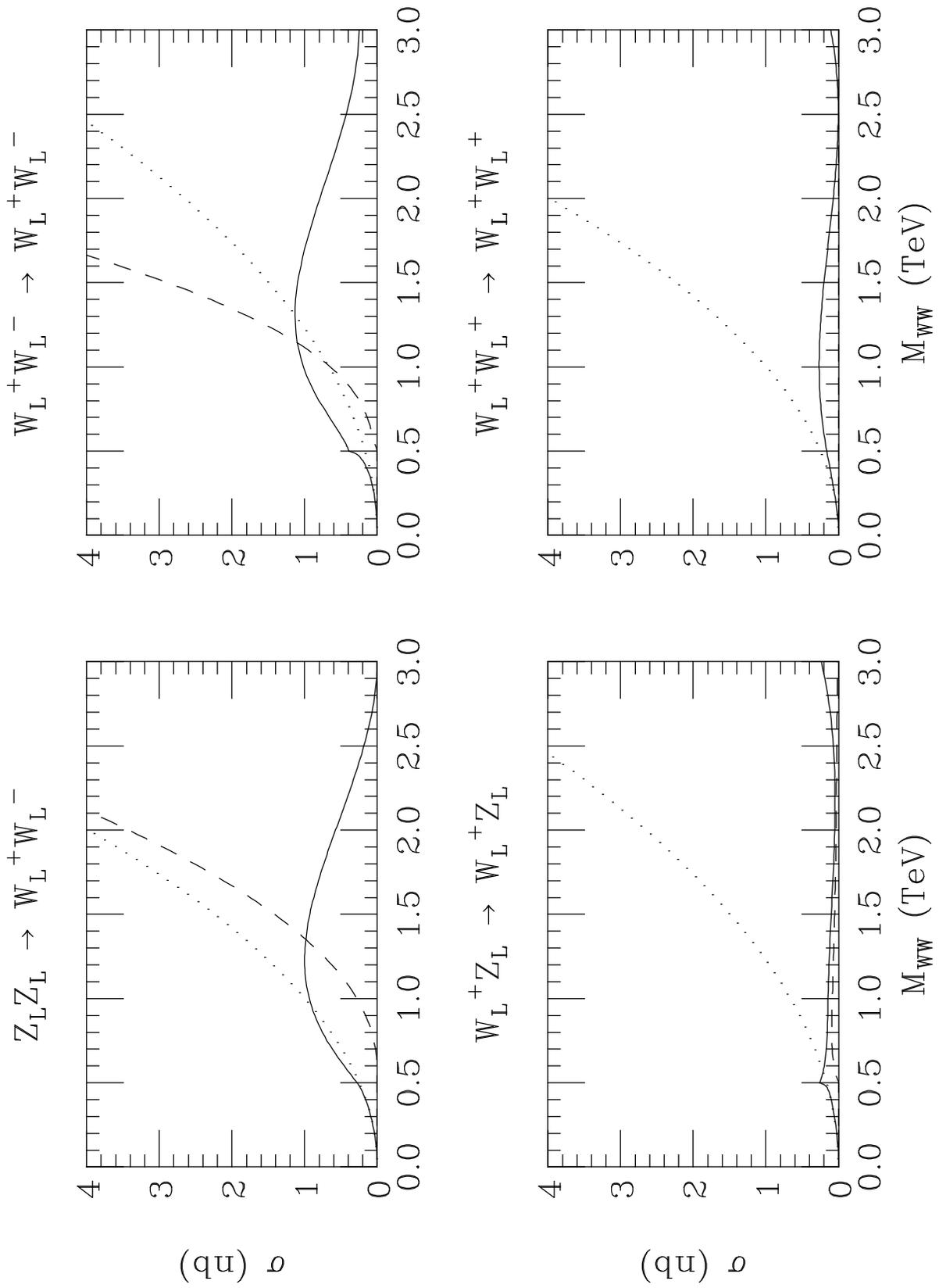
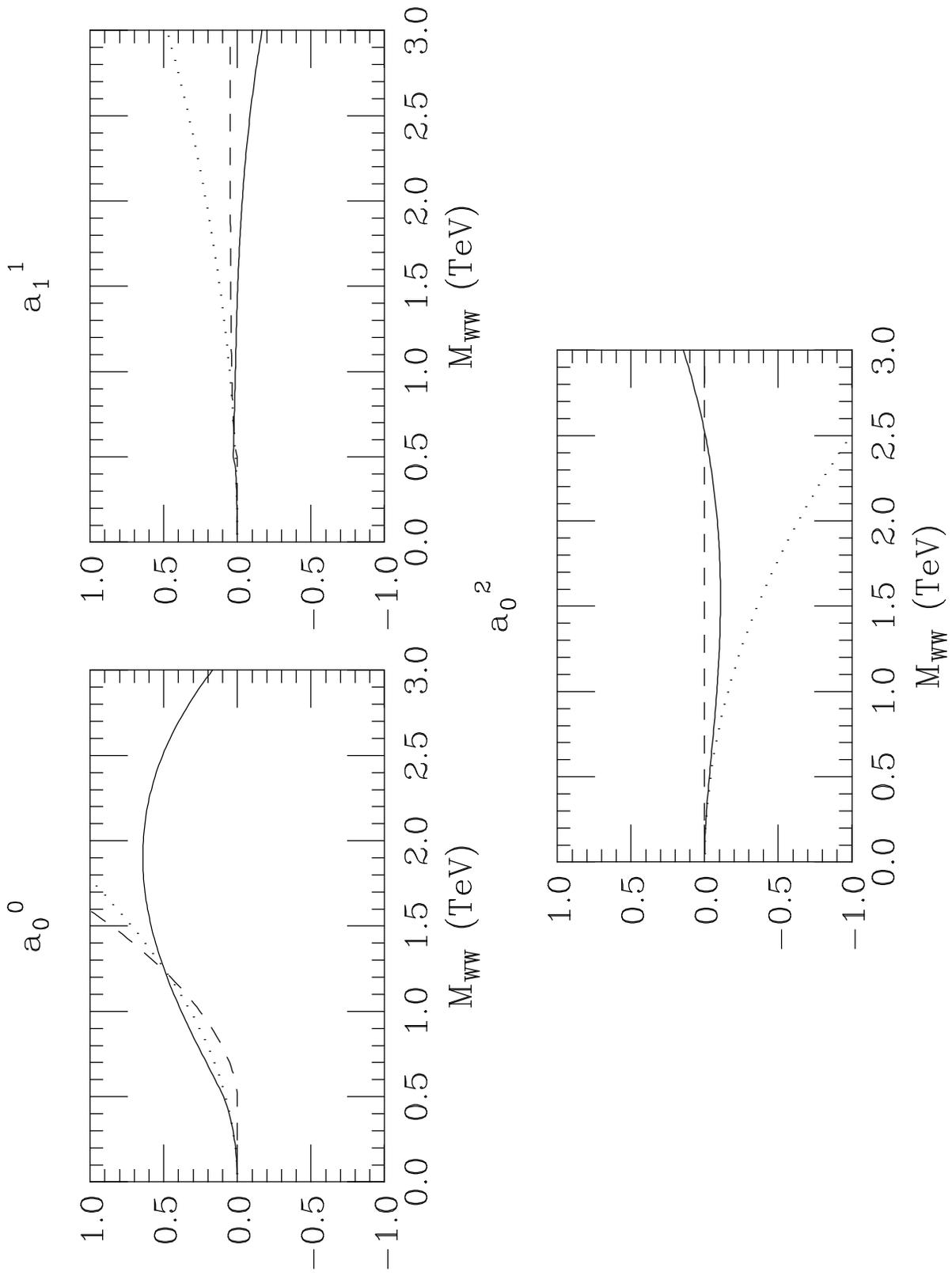
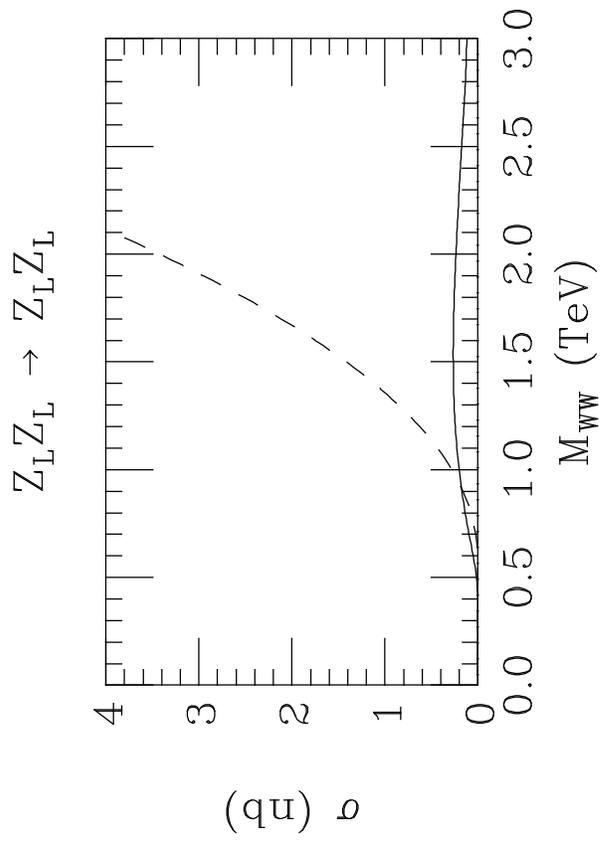
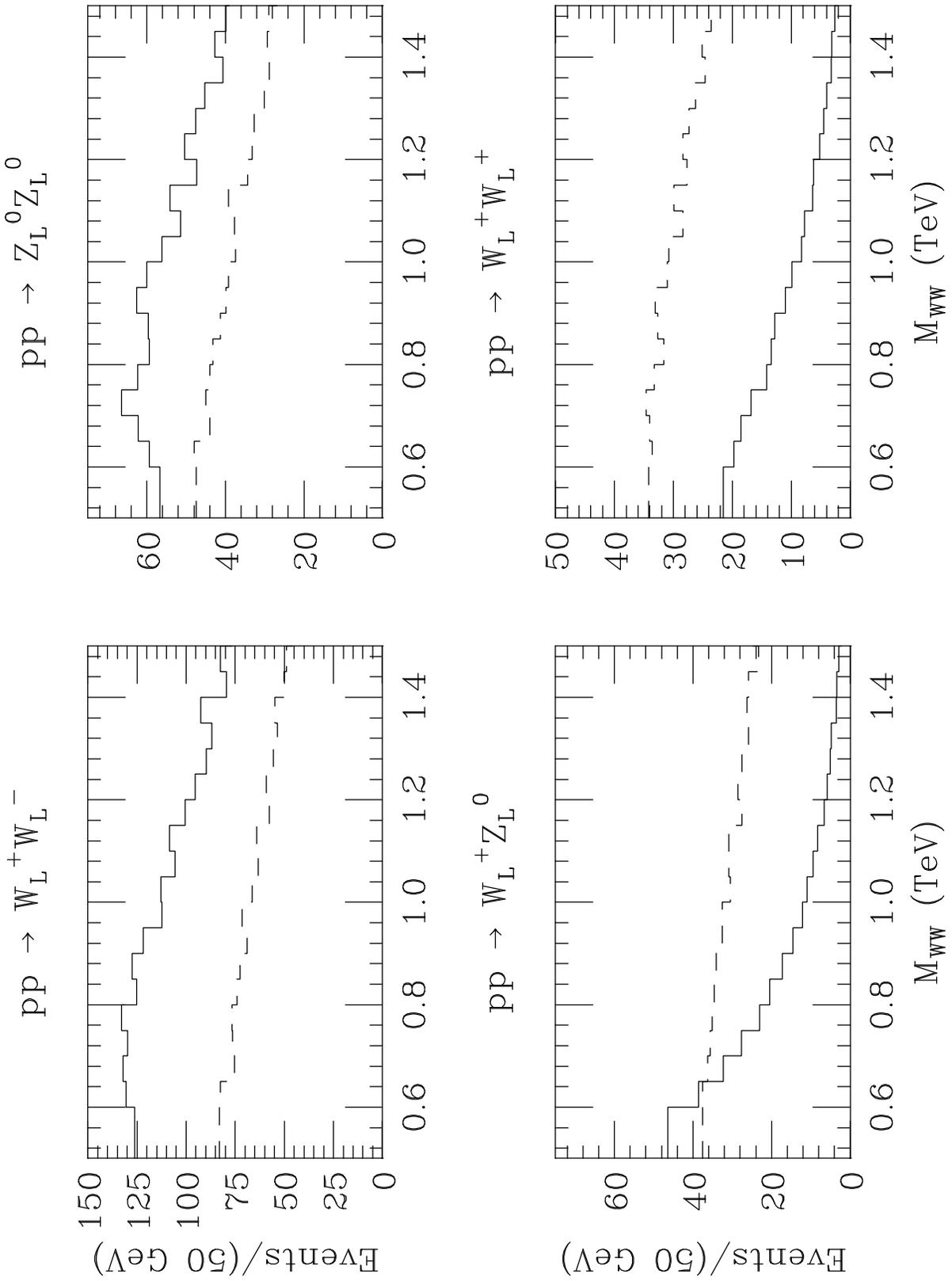
\end